\documentclass[12pt]{article}
\usepackage{times}
\usepackage{geometry}
\usepackage{graphicx}
\usepackage{parskip}
\usepackage[utf8]{inputenc}
\usepackage{enumitem,amssymb}
\usepackage{ragged2e}
\newlist{thematic}{itemize}{8}
\setlist[thematic]{label=$\square$}
\usepackage{pifont}
\newcommand{\cmark}{\ding{51}}%
\newcommand{\done}{\rlap{$\square$}{\raisebox{2pt}{\large\hspace{1pt}\cmark}}%
\hspace{-2.5pt}}

\usepackage{natbib}
\bibpunct{(}{)}{;}{a}{}{,}

\usepackage{natbib}
\bibpunct{(}{)}{;}{a}{}{,}

\usepackage{paralist}
\usepackage{sidecap,wrapfig}
\usepackage{multicol}
\usepackage[pdftex]{thumbpdf}
\usepackage{relsize}
\usepackage{mathptmx}
\usepackage{enumitem}
\usepackage{epstopdf}
\DeclareGraphicsExtensions{.pdf,.png,.jpg,.jpeg,.gif}
\def\simlt{\mathrel{\rlap{\lower 3pt\hbox{$\sim$}}\raise 2.0pt\hbox{$<$}}}
\def\simgt{\mathrel{\rlap{\lower 3pt\hbox{$\sim$}} \raise 2.0pt\hbox{$>$}}}
\newcommand{\lta}{\mathrel{\spose{\lower 3pt\hbox{$\mathchar''218$}}
      \raise 2.0pt\hbox{$\mathchar''13C$}}}
\newcommand{\gta}{\mathrel{\spose{\lower 3pt\hbox{$\mathchar''218$}}
      \raise 2.0pt\hbox{$\mathchar''13E$}}}

\def\farcs{%
 \mbox{%
  \kern  0.13ex.%
  \kern -0.95ex\raisebox{.9ex}{\scriptsize$\prime\prime$}%
  \kern -0.1ex%
 }%
}%

\def\lesssim{\mathrel{\hbox{\rlap{\hbox{\lower3pt\hbox{$\sim$}}}\hbox{\raise2pt\hbox{$<$}}}}}
\def\gtrsim{\mathrel{\hbox{\rlap{\hbox{\lower3pt\hbox{$\sim$}}}\hbox{\raise2pt\hbox{$>$}}}}}

\newcommand{\beq}{
\begin{equation}
}
\newcommand{\eeq}{
\end{equation}
}
\newcommand{\beqa}{
\begin{eqnarray}
}
\newcommand{\eeqa}{
\end{eqnarray}
}

\makeatletter
\providecommand{\ion}[2]{#1$\;$\textsmaller{\@Roman{#2}}}
\makeatother

\newcommand{\msun}     {\ensuremath{{M}_{\scriptscriptstyle \odot}}}
\newcommand{\lsun}{\ensuremath{{L}_{\scriptscriptstyle \odot}}}

\newcommand{\kms}      {\ensuremath{~\mathrm{km~s^{-1}}}}

\newcommand{\ergs}     {\ensuremath{~\mathrm{erg\,s^{-1}}}}

\newcommand{\mbh}      {\ensuremath{M}}

\begin{document}
\raggedright
\huge
Astro2020 Science White Paper \linebreak

Black Holes Across Cosmic Time \linebreak
\normalsize

\noindent \textbf{Thematic Areas:} \hspace*{60pt} $\square$ Planetary Systems \hspace*{10pt} $\square$ Star and Planet Formation \hspace*{20pt}\linebreak
$\square$ Formation and Evolution of Compact Objects \hspace*{31pt} $\square$ Cosmology and Fundamental Physics \linebreak
  $\square$  Stars and Stellar Evolution \hspace*{1pt} $\square$ Resolved Stellar Populations and their Environments \hspace*{40pt} \linebreak
  \done    Galaxy Evolution   \hspace*{45pt} $\square$             Multi-Messenger Astronomy and Astrophysics \hspace*{65pt} \linebreak
  
\textbf{Principal Author:}

Name:	Kayhan G{\"u}ltekin
 \linebreak						
Institution:  Dept.\ of Astronomy, University of Michigan
 \linebreak
Email: kayhan@umich.edu
 \linebreak
Phone:  +1 734.255.7082
 \linebreak
 
\textbf{Co-authors:} (names and institutions)
  \linebreak
Aaron Barth (University of California, Irvine), 
Karl Gebhardt (University of Texas), 
Jenny Greene (Princeton University), 
Luis Ho (Kavli Institute for Astronomy and Astrophysics, Peking University), 
St{\'e}phanie Juneau (National Optical Astronomy Observatory),
Chung-Pei Ma (University of California, Berkeley),
Anil Seth (University of Utah),
Vivian U (University of California, Irvine), 
Monica Valluri (University of Michigan), 
Jonelle Walsh (Texas A\&M)

\textbf{Abstract  (optional):}

Supermassive black holes are located at the center of most, if not all, massive galaxies. They follow close correlations with global properties of their host galaxies (scaling relations), and are thought to play a crucial role in galaxy evolution. Yet, we lack a complete understanding of fundamental aspects of their growth across cosmic time.  In particular, we still do not understand: (1) whether black holes or their host galaxies grow faster and (2) what is the maximum mass that black holes can reach. The high angular resolution capability and sensitivity of 30-m class telescopes will revolutionize our understanding of the extreme end of the black hole and galaxy mass scale. With such facilities, we will be able to dynamically measure masses of the largest black holes and characterize galaxy properties out to redshift $z\sim1.5$.  Together with the evolution of black hole-galaxy scaling relations since $z\sim1.5$, the maximum mass black hole will shed light on the main channels of black hole growth. 

\pagebreak
\section{The Largest Black Holes}

Almost all ellipticals and large galaxy bulges contain supermassive black holes.  It is well known that in the local universe the mass of black holes is tightly correlated to the large-scale properties of the stellar spheroid of the host galaxy (e.g., mass, luminosity, and stellar velocity dispersion).  These ``scaling relations'' strongly suggest a fundamental link between the evolution of a galaxy and its central black hole, tying together quantities of very different length- and mass-scales.  The discovery of the scaling relations in the local universe has profoundly affected our understanding of galaxy evolution in general and affected the study of the evolution and luminosity function of massive ellipticals in particular.  Theoretical ideas for the underlying causes of the scaling relations include self-regulated feedback originating from galaxy-merger-triggered active galactic nuclei (AGN) or star formation.

It was high-spatial resolution spectroscopy\,---\,starting in earnest with \emph{Hubble Space Telescope} and then progressing to adaptive optics (AO) on 10-m class telescopes\,---\,that enabled the measurement of black hole masses, the discovery of the empirical scaling relations, and thus this advance in understanding.  Despite this advance, there is still no definitive answer as to what has led to the tight scaling relations in the local universe, and recent measurements show evidence of large scatter in the highest and lowest mass regimes.  Two key observational ingredients are missing: demographics of the largest black holes and black hole--galaxy scaling relations as a function of redshift.  With 10 mas angular resolution spectroscopy enabled by 30-m class extremely large telescopes (ELTs), it will be possible to do both.

\subsection{What is the maximum black hole mass?}
Demography of the largest black holes in particular has been relatively poorly explored to this point despite the importance of the role that these black holes are thought to play.  
In the local universe, the two largest black holes from direct dynamical measurements have $M=1.5$--$2 \times 10^{10}\ \msun$ \citep{2011Natur.480..215M, 2016Natur.532..340T}.
Indirect evidence points to the existence of black holes perhaps more massive than 
these, though this is based on AGN emission properties, which are less reliable than dynamical mass estimates.  For example, the most luminous quasars reach 
$R$-band luminosities of $4 \times 10^{48}\ \ergs$ ($10^{15}\ \lsun$) and therefore an 
Eddington-limit-based mass of $3 \times 10^{10}\,\msun$.  
\citet{2009MNRAS.393..838N} found that Eddington-limited growth implies an 
upper limit to black hole masses at every redshift and corresponds to
$M\sim10^{10}\,\msun$ at $z=0$.  \citet{2009MNRAS.399L..24G} found a $\mbh = 4
\times 10^{10}\,\msun$ black hole based on SED models including hard
X-ray observations with \emph{Swift} in a $z=3.37$ blazar, and in a
similar study, \citet{2010MNRAS.405..387G} found 7 AGNs
with $\mbh > 3 \times 10^{9}$ at $z > 2$.   
\textbf{With ELTs, it will be possible to dramatically expand the population of the largest black holes with dynamical measurements in the local universe \citep{2013ApJ...764..184M,  2013ARA&A..51..511K, 2016ApJ...818...47S} and up to $z=1.5$.}

\subsection{The importance of finding the most massive black holes}
The largest black holes play an important role in a number of areas of astrophysics.  Although the largest black holes are rare, they still play a critical role in black hole demographics.  The upper end of the black hole mass function has a strong influence on the integrated black hole mass density.  Additionally, the most massive black holes shape the upper end of the galaxy mass function.

Another area of importance of large black holes is the major role they are thought to play in regulating the gaseous ecosystem of galaxy clusters.  It is hypothesized that cluster cooling flows are inhibited 
by AGN heating from central galaxy.  \emph{Chandra} observations support a picture in 
which episodic AGN outbursts in brightest cluster galaxies may heat the
intracluster medium \citep{2005ApJ...634..955V}; the energetics required to
terminate cooling flows implies $\mbh > 10^{10}\;\msun$ for many
clusters \citep{2002MNRAS.335L..71F}.  Even after taking into account other
sources of heating such as thermal conduction, dynamical friction, and
supernovae, the most massive clusters still require the largest black
holes to quench cooling.  
A related issue is that in order to produce the sharp cutoff in quenching at halo masses $M \sim 10^{12}\ \msun$, non-linear black hole growth is required \citep{2017MNRAS.465...32B}.  Thus getting precise black hole masses is important in understanding AGN feedback at the largest scales.

The largest black holes in the largest galaxies can also provide insight 
into the connection between galaxy mergers and black hole mergers.  
The flat luminosity cores in massive ellipticals are one of their most
distinctive features.  It has been argued that the cores are scoured
out by the inspiral of black holes during galaxy mergers
\citep{1980Natur.287..307B, 1997AJ....114.1771F, 2003ApJ...596..860M,
2007ApJ...662..808L}. 
The inspiral of black holes will lead to binary black holes, which can eventually emit gravitational waves observable by pulsar timing arrays.
While these processes can happen in many galaxies, it is most prominent in the largest galaxies, the most likely merger remnant candidates.

\section{Black Hole Growth over Cosmic Time}
\begin{figure}[htb!]
\begin{center}
\includegraphics[width=\textwidth]{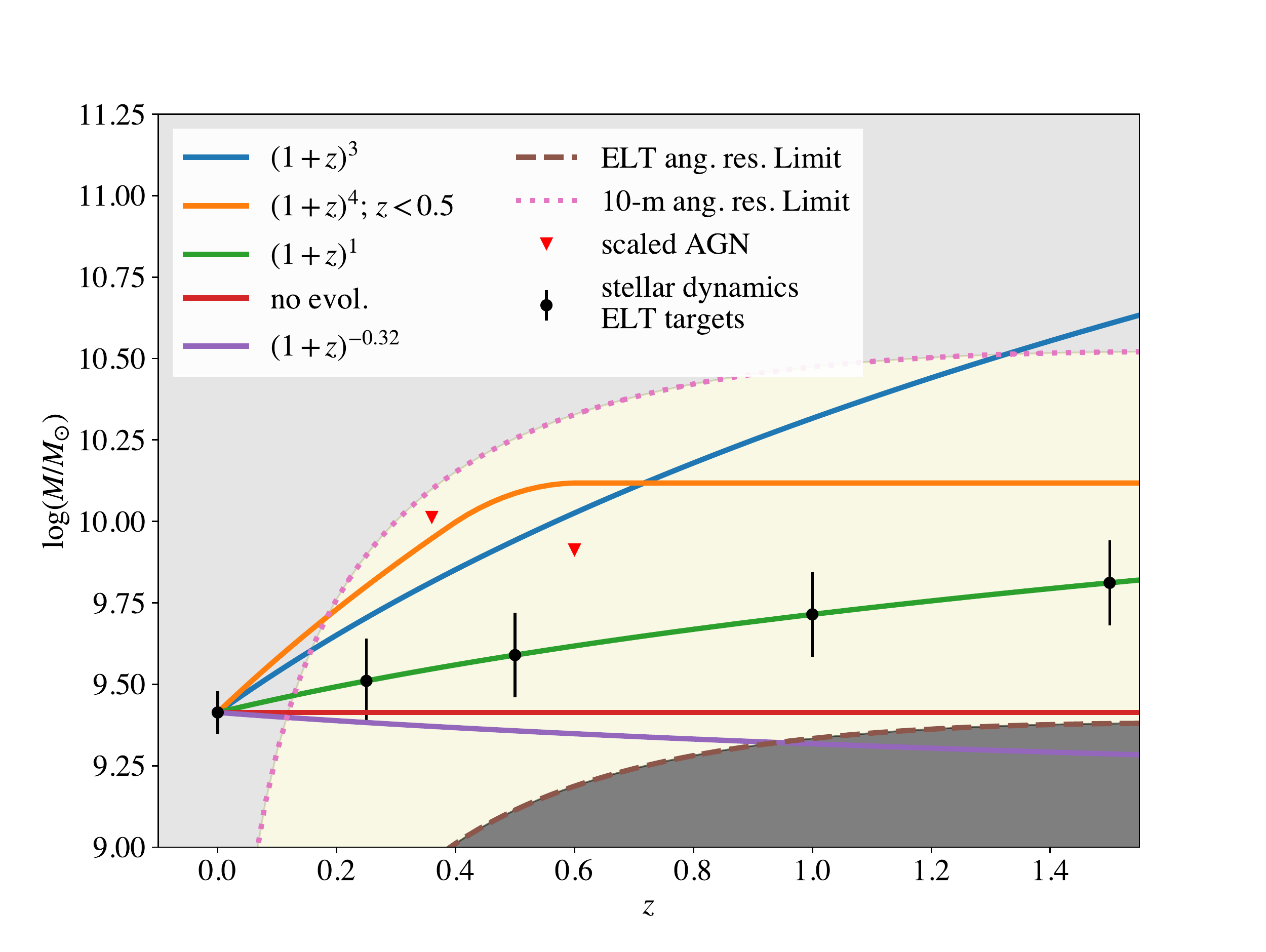}
\caption{Plot of mean logarithmic black hole mass as a function of redshift for galaxies with velocity dispersion $\sigma = 325\,\mathrm{km\,s^{-1}}$. The solid lines show several phenomenological models and predictions from simulations for the evolution of black hole mass scaling relations in large galaxies.  Models with positive slope indicate that black hole growth precedes galaxy growth; those with negative slope indicate that galaxy grows first; and those with zero or near zero slope come from near equilibrium co-evolution for $z<2$.  Red triangles show data from lower-mass AGN extrapolated to large black holes under the assumption that the ratio of black hole to galaxy mass is the same at a given redshift. The black circles show the uncertainties with which we can test these models with 40 sources at $z=0$ and 10 sources in each redshift bin $z > 0$, which is achievable with an ELT large program.  The data points are arbitrarily put on the $(1 + z)^1$ growth curve to demonstrate the ability to test models.  The dotted and dashed lines show the minimum detectable mass based only on angular resolution for a 10-m class telescope and ELT, respectively.  In addition the angular resolution limit, the surface brightness dimming of sources makes them prohibitively expensive for 10-m class telescopes but achievable for ELTs.  \emph{Measuring black hole masses at cosmological distances is only possible with diffraction-limited ELTs and opens up new discovery space.} }
\label{f:mvsz}
\end{center}
\end{figure}

How black hole scaling relations evolve with redshift is unknown at this time, even though significant effort has been put into this question \citep[e.g.,][]{2019arXiv190300003Y}. The existence of the black hole scaling relations may be a natural consequence of black hole growth and feeding processes as galaxies evolve, but this relationship may not hold in a younger universe when galaxy properties and their environmental conditions were different.  
Figure \ref{f:mvsz} shows models of the mean logarithmic mass of a black hole in a host galaxy with stellar velocity dispersion $\sigma = 325\,\kms$ as a function of redshift.  Models with a  positive slope show that the black hole lags behind the galaxy in terms of mass evolution, whereas models with  negative slopes show that black holes grow close to their full mass at early times and thus look ``overmassive'' for their host galaxy at high redshift.  Finally, models with flat slopes indicate that black holes and their host galaxies evolve in lockstep with neither deviating from the $z=0$ relation at any redshift.  It is clear that with the angular resolution of current 10-m class telescopes, dynamical black hole mass measurement outside of the local universe is impossible.

\emph{Analytical arguments, semi-analytical models, and simulations disagree on the fundamental question of whether black holes grow to their $z=0$ masses before, after, or along with their host galaxy.}
For example, based on a simple analytic model where the black hole mass is governed by the gravitational binding energy of its host halo, \citet{2011MNRAS.413.1158B}  predicted galaxy growth to precede black hole growth. \citet{2009ApJ...707L.184J} come to similar conclusions based on simulations of mergers and the inconsistency of overmassive black holes with current scaling relations.  On the other hand, some theoretical work predicts black holes to grow first at high redshift based on consideration of galaxy mergers \citep{2006MNRAS.369.1808C} or cosmological simulations \citep{2006MNRAS.370..645B}.  Finally, \citet{2004ApJ...600..580G} find theoretical support for lockstep evolution based on supernova and AGN feedback modeling.

There is a clear need for an observational breakthrough to inform theory as the current observational constraints of black hole scaling evolution are similarly inconsistent.  Based on AGN scaling relations, various groups have argued for dramatic evolution, with black holes' being overly massive relative to present day relations at moderate redshift \citep[$z \sim 0.3$--$0.5$;][]{2006ApJ...645..900W, 2008ApJ...681..925W, 2007ApJ...667..117T, 2010ApJ...708.1507B, 2015ApJ...799..164P},  intermediate \citep[$z \sim 2$;][]{2006NewAR..50..782M}, and high redshift \citep[$z \sim 6$;][]{2016ApJ...816...37V}, while other groups have argued that these offsets can be completely ascribed to various observational biases \citep{2010ApJ...708..137M, 2012AAS...21943503S, 2013ApJ...764...80S, 2015ApJ...799..173S, 2015ApJ...805...96S}   These observational constraints are all estimated from AGN emission properties, which are fundamentally less secure than direct dynamical measurements. Only direct dynamical constraints can resolve this debate and chart the coevolution (or not) of black holes and galaxies.

\textbf{ELTs will be able to measure the masses of black holes up to $z=1.5$ using stellar dynamical techniques (and up to $z=2.0$ with gas dynamical techniques) that are largely immune to the biases above and will let us distinguish among the different coevolution models, yielding physical insights into the details of black hole growth and the relative influence of black holes and galaxies on each other.}

\section{Observing Large Black Holes with ELTs in the 2020's}

Whether in the local universe or at high redshift, the ability to measure black hole mass with stellar dynamical models requires angular resolution sufficient to resolve the black hole's sphere of influence, which has radius $r_{\mathrm{infl}} = G M \sigma^{-2}$, where $\sigma$ is the host galaxy stellar velocity dispersion.  Because black hole masses are highly correlated with $\sigma$, for a given angular resolution ($\theta_{\mathrm{res}}$), the minimum black hole mass that can be realistically detected depends only on the angular diameter distance, $D_A$.  For an $M$--$\sigma$ relation of the form
%
    $M = M_0 (\sigma / \sigma_0)^{\beta}$,
%
the minimum detectable mass is 
\begin{equation}
    M_{\mathrm{min}} = \left[ \frac{\theta_{\mathrm{res}}}{G M_0^{2/\beta}} \sigma_0^2 D_A \right]^{\beta / (\beta - 2)}.
\end{equation}
Using the $M$--$\sigma$ relation from, e.g., Kormendy \& Ho (2013), $M_0 = 3.09 \times 10^{9}\,\msun$, $\sigma_0 = 200\,\kms$, and $\beta = 4.28$.  Thus $M_{\mathrm{min}} = 8.6 \times 10^{6}\,\msun (\theta_{\mathrm{res}} D_A)^{1.88}$ for $\theta_{\mathrm{res}}$ in arcsec and  $D_A$ in Mpc.  Fig.\ \ref{f:mvsz} shows $M_{\mathrm{min}}(z)$ for 10-m telescopes and ELTs. Thus the improvement in angular resolution afforded by ELTs opens up a large discovery space and allows us to measure black hole masses at cosmological distances.

In addition to angular resolution, a large collecting area is necessary to make spectroscopic observations of high-$z$ galaxies at high enough signal to noise to be useful.  The largest black holes are found in the largest galaxies, which typically have low surface brightness (at $z=0$).  As surface brightness dims as $(1 + z)^4$, integration times for a fixed surface-brightness become expensive.  This is mitigated by the fact that central surface brightness of galaxies at fixed velocity dispersion brightens with redshift as $(1 + z)^{1.5}$ \citep{2003ApJ...597..239G} so that it is achievable with ELTs out to $z=1.5$.  At higher redshifts ($z=2$), it will be possible to measure black hole masses with ionized gas-dynamical mass measurements once suitable targets with regularly rotating ionized gas disks can be identified.

To make use of the angular resolution for stellar dynamical measurements of black hole masses, sufficiently good integral field spectroscopy (IFS) is also needed.  Spectral resolution of $R = 4000$--$6000$ with a 10 mas (or better) pixel scale with a field of view of 0.2--0.6 arcsec$^2$ will do the job.  Though $R\sim5000$ is more than what is required for the largest black holes, such spectral resolution will be needed for complementary studies of smaller black holes in the local universe, which can be efficiently observed.  For larger black holes, binning in spectral resolution can be done to increase signal to noise.  An ELT with this kind of instrumentation can be used to map out the stellar kinematics using the 2.3 $\mu$m and the 1.6 $\mu$m CO bandheads for $z < 0.5$ and the 0.85 $\mu$m Ca II triplet lines at higher redshifts.  

The final component of measuring a black hole mass is detailed modeling of the surface brightness and kinematics.  This is a non-trivial analysis that takes significant computational resources for each galaxy modeled.  To make the most of the data from ELTs, we will need the best modeling codes.  There is an argument to be made in favor of a shared, open modeling code that is available to the community. Having a shared infrastructure would make the science more accessible to a wider group of scientists, enhance the long-term value of archival ELT data, and improve robustness and reproducibility of results.

\section{Recommendations}
Understanding the evolution of the largest black holes across cosmic time requires investment in and exploitation of high-angular resolution integral field spectroscopy on a 30-m class telescope behind adaptive optics.  In order to measure the largest black holes in the nearby universe as well as to measure the evolution of black-hole--galaxy scaling relations over cosmic time, angular resolution of 10 mas or better is needed with IFS. 
In addition to the instrumentation needed, robust modeling codes are needed to make the most of the data.
As has been learned over the past several decades, evolution of black holes and galaxies can no longer be studied in isolation.  Key observational advances to advance the state of these fields will become accessible in the 2020s from direct measurement of the largest black holes across cosmic time.

\pagebreak
\textbf{References}

\bibliographystyle{apj_short_prop.bst}
\bibliography{smbhastro2020bib}

\end{document}